\documentclass[dvipsnames,aps,pre,nobibnotes,twocolumn,superscriptaddress,nolongbibliography]{revtex4-2}

\usepackage{graphicx}

\usepackage{float}
\usepackage{enumitem}
    \newlist{ipclist}{enumerate}{1}
    \setlist[ipclist,1]{label=\bf\Alph*}
\usepackage[x11names, table, svgnames]{xcolor}
    \definecolor{hypercitecolor}{HTML}{4287f5}
\usepackage[T1]{fontenc}
\usepackage[latin9]{inputenc}
\usepackage[english]{babel}
\usepackage{mathtools}
\usepackage{amsthm}

\usepackage{amssymb}
\usepackage{array}
\usepackage{xfrac}
\usepackage{mathrsfs}
\usepackage{bm}
\usepackage{xparse}
\usepackage{siunitx}
\usepackage{stackrel}
\usepackage{nicefrac}
\usepackage{braket}
\usepackage[acronym]{glossaries}
    \newacronym{lrg}{LRG}{Laplacian Renormalization Group}
    \newacronym{sl}{SL}{Signed Laplacian}
    \newacronym{si}{SI}{Supplementary Information}
    \newacronym[\glslongpluralkey={Spin Glasses}]{sg}{SG}{Spin Glass}
    \newacronym{rbim}{RBIM}{Random Bond Ising Model}
    \newacronym{rsb}{RSB}{Replica Symmetry Breaking}
    \newacronym{ea}{EA}{Edwards--Anderson}
    \newacronym{gc}{GC}{Giant Component}
    \newacronym{gsd}{GSD}{Graph-Spectral Decomposition}
    \newacronym{mse}{MSE}{Mean Squared Error}
    \newacronym{mc}{MC}{Monte Carlo}

\let\transp\intercal
\let\vec\bm
\let\adjacencymatrixsymb A
\let\bigOnotationsymb O
\let\couplingmatrixsymb J
\let\critsymb c
\let\degreematsymb D

\let\edgesetsymb E
\let\entropysymb S

\let\graphsymb G

\let\hamiltoniansymb H
\let\laplaciansymb L
\let\nodedegreesymb k
\let\partitionfunctionsymb Z
\let\probabilitysymb P

\let\sinkoperatorsymb S

\let\vertexsymb V
\let\weightedadjacencymatrixsymb W

\newcommand{\signshrt}{sign}

\newcommand{\adj}{\adjacencymatrixsymb}

\newcommand{\lapl}{\laplaciansymb}

\let\slapl\Slapl

\DeclareMathOperator{\sign}{\signshrt}

\DeclarePairedDelimiter{\abs}{|}{|}

\NewDocumentCommand{\dd}{s o m}{
  \IfBooleanTF{#1} 
    {\mathrm{d}#3} 
    {\IfNoValueTF{#2} 
      {\mathrm{d}#3} 
      {\mathrm{d}^{#2}#3} 
    }
}

\NewDocumentCommand{\dv}{s o m m}{
  \IfBooleanTF{#1} 
    {{\mathrm{d}#3}/{\mathrm{d}#4}} 
    {\IfNoValueTF{#2} 
      {\frac{\mathrm{d}#3}{\mathrm{d}#4}} 
      {\frac{\mathrm{d}^{#2}#3}{\mathrm{d}#4^{#2}}} 
    }
}

\NewDocumentCommand{\pdv}{s o m m}{
  \IfBooleanTF{#1} 
    {{\partial #3}/{\partial #4}} 
    {\IfNoValueTF{#2} 
      {\frac{\partial #3}{\partial #4}} 
      {\frac{\partial^{#2} #3}{\partial #4^{#2}}} 
    }
}
\usepackage[caption=false]{subfig}
\usepackage{tikz}
    \usetikzlibrary{positioning}
    \usetikzlibrary{fit}
    \tikzset{graph node/.style={draw, circle}}
    \tikzset{dgraph node/.style={draw, densely dashed, circle}}

\PassOptionsToPackage{hyphens}{url}
\usepackage{hyperref}
    \hypersetup{
        colorlinks,
        citecolor=blue,
        urlcolor=blue,
        linkcolor=blue,
    }

\usepackage{algorithm}
\usepackage{algpseudocode}
\usepackage{amsmath}
\makeatletter
    \renewcommand\paragraph{%
      \@startsection{paragraph}{4}{\z@}%
        {3.25ex \@plus1ex \@minus .2ex}%
        {-1em}%
        {\normalfont\normalsize\bfseries}%
    }
    \newcommand{\phantomlabel}[2]{
        \protected@write\@auxout{}{
            \string\newlabel{#2}{
                {\@currentlabel#1}{\thepage}
                {\@currentlabel#1}{#2}{}
            }
        }
        \hypertarget{#2}{}
    }
    \NewDocumentCommand{\FigRef}{m}{%
      \ifvmode Figure~\ref{#1}%
      \else Fig.~\ref{#1}%
      \fi
    }
    \renewcommand{\eqref}[1]{Eq.~(\ref{#1})}
    
\makeatother

\usepackage{pdfpages}
\makeatletter
\AtBeginDocument{\let\LS@rot\@undefined}
\makeatother

\begin{document}
\title{Chladni States in Ising Spin Lattices}
\author{Giulio Iannelli}
    \email{giulio.iannelli@cref.it}
    \affiliation{`Enrico Fermi' Research Center (CREF), Via Panisperna 89A, 00184 - Rome, Italy}
    \affiliation{Dipartimento di Fisica, Universit\`a degli Studi di Palermo, 90133 Palermo, Italy}
\author{Pablo Villegas}
    \email{pablo.villegas@cref.it}
    \affiliation{`Enrico Fermi' Research Center (CREF), Via Panisperna 89A, 00184 - Rome, Italy}
    \affiliation{Instituto Carlos I de F\'isica Te\'orica y Computacional, Univ. de Granada, E-18071, Granada, Spain.}

\begin{abstract}
Low-temperature spin dynamics can become trapped in long-lived patterns shaped by the geometry of the interaction network. We introduce \emph{Chladni states}, the binary spin configurations obtained by applying the sign function to Laplacian eigenmodes, and show that their Ising energy collapses onto a size-independent spectral curve in ferromagnets and frustrated antiferromagnets. This graph-spectral decomposition reconstructs frozen configurations, tracks memory of initial conditions, and detects, through the number of modes it requires, the loss of mesoscopic order in frustrated and spin-glass regimes. Finally, in spin-glass instances, a restricted search over low signed-Laplacian modes reaches energies comparable to those of simulated annealing while exploring a space of far smaller dimension.
\end{abstract}

\maketitle 

Chladni plates are metal surfaces covered with sand and subjected to vibrations \cite{Chladni1802}. At specific frequencies, standing waves form, producing distinct sand patterns on the plate. As an early technique for visualizing mechanical vibrations, Chladni plates show how the shape, geometry, and symmetries of a substrate constrain the resulting modes \cite{Sudhir2017}. These ideas have inspired applications in topological mechanical metamaterials \cite{Zhou2016, Kopitca2021, Latifi2019}, where geometric design principles guide wave localization and energy transport. An analogous selection occurs in non-ergodic spin systems, where the geometry of the interaction network determines which metastable patterns can be sustained.

In the context of Ising ferromagnets, a sudden quench from high to zero temperature often traps the system in metastable domain mosaics rather than allowing relaxation to the globally ordered state \cite{Ramgopal2022, Olejarz2012}. These configurations can persist indefinitely, with relaxation times diverging with system size in the low-temperature phase, a hallmark of broken ergodicity \cite{Bray1994, Cirillo1998}. Furthermore, the final configuration depends sensitively on the underlying substrate and microscopic details \cite{Nattermann1997}, leading to a rich mosaic of geometric patterns \cite{Redner2009, Redner2001}. This produces a complex landscape of frozen configurations whose organization remains only partially understood~\cite{Olejarz2012}.

Frustration introduces a further layer of complexity. For example, in antiferromagnets on triangular lattices, not all bonds can be satisfied simultaneously due to geometric incompatibilities, in contrast to square lattices \cite{Ramirez2003, Anderson1956, Wannier1950, Saito1984}. The result is a highly degenerate manifold of low-energy states with slow relaxation dynamics and large residual entropy \cite{Ramirez2003, Skjaervo2020}. A central challenge in these frustrated systems is both tracking the system's trajectory and characterizing the landscape of metastable configurations, as extensive degeneracy and kinetic constraints generate slow relaxation and long-lived non-equilibrium states \cite{Skjaervo2020}. Remarkably, recent advances have enabled the macroscopic manipulation of such magnetic patterns, offering a framework for designing magnetic memory devices via controlled frustrated spin states \cite{Wang2025}.

\glspl{sg}, the paradigm of disordered magnetic systems with competing interactions among constituent spins \cite{Edwards1975}, are another archetypal example of the ground-state degeneracy problem \cite{Kirkpatrick1977, Hubert}. There, the proliferation of nearly degenerate states makes the low-temperature behavior strongly history-dependent \cite{Bray1987, Parisi1979, Castellani2005}, so that thermal averages no longer determine the configuration actually reached \cite{Bray1987, Jonason1998}. However, recent studies suggest that the \gls{sg} phase may have a deep graph-geometric origin \cite{Iannelli2025}, opening the door to addressing the problem from a spectral perspective. Still, deciphering the nature of low-temperature \glspl{sg} in short-range models, i.e., determining the number of pure states, reconstructing ground states \cite{Fan2023}, and understanding correlated disorder \cite{Mezard2024}, remains a major open problem in the statistical mechanics of disordered systems \cite{Newman2004}.

Here, we present a graph-spectral framework linking metastable spin states to the eigenmodes of the Laplacian operator defined by the interaction graph. By binarizing these eigenvectors, we obtain a distinguished family of discrete configurations, which we call Chladni states, and we characterize their energy, their dynamical stability, and their relation to the degenerate ground-state manifold in ferromagnetic, antiferromagnetic, and \gls{sg} systems at low
temperature. In this view, the spectrum of the interaction graph determines on its own which frozen configurations a lattice can sustain, and which of them the dynamics cannot leave. As a proof of concept, we also explore graph-spectral
compression and classification schemes, outlining possible applications to structured magnetic memories and physics-inspired information processing.

\newpage
\section{Chladni states}
Any Ising spin lattice can be described by the Hamiltonian
\vspace{-0.25em} 
\begin{equation}
    \mathcal{H}=-\sum_{\langle ij\rangle}J_{ij}\sigma_i\sigma_j,
    \label{Ising}
\end{equation}
where $\sigma_i\in\{\pm1\}$ is the spin variable at site \(i\) and the sum runs over distinct pairs. The interaction matrix \(J_{ij}\) encodes both the geometry of the lattice and the nature of the couplings: \(J_{ij}\geq0\) for ferromagnetic interactions, \(J_{ij}\leq0\) for antiferromagnetic interactions, and mixed signs for random-bond or \gls{sg} systems. In the latter case, frustration and degeneracy depend on the distribution of positive and negative couplings and on the microscopic structure of the underlying lattice \cite{Iannelli2025}.

For ferromagnets, the standard field-theoretical description of \eqref{Ising} introduces a coarse-grained magnetization field, $\phi(\mathbf{x})$, obtained by averaging over local spins. This leads to the Landau--Ginzburg Hamiltonian, $\mathcal{H}=\beta^{-1}\int \mathrm{d}^d\mathbf{x}\,[\alpha^2(\nabla\phi)^2+\mu^2\phi^2+\lambda\phi^4+\mathcal{O}(\phi^6)]$, where $\beta$ is the inverse temperature, and $\alpha$, $\mu$, and $\lambda$ are phenomenological parameters depending on microscopic interactions \cite{Kardar,Binney,Amit}. This continuum formulation underlies the perturbative renormalization-group description of critical phenomena: beyond a mesoscopic scale, lattice details become irrelevant, and universal properties are controlled mainly by symmetries, dimensionality, and interaction range. However, this abstraction can obscure an important fact: the macroscopic behavior of real magnetic materials is often shaped by their discrete crystalline structure, defects, and microstructure~\cite{Simonov2020}, which can play a decisive role in determining their collective phases~\cite{Hubert, Hehn1996}. Similar geometry-dependent effects arise in ferroelectrics, perovskites, and other functional materials~\cite{Nahas2020, Ramesh2019, Falsi2024}.

In discrete spin models, universality still holds: as long as microscopic symmetries and interaction ranges are preserved, the details of the lattice tiling become irrelevant at sufficiently large scales~\cite{MarroBook}. Here, however, we focus on the complementary question of how metastable configurations retain signatures of the discrete interaction geometry before this information is lost under coarse-graining or continuum descriptions. We therefore consider the analysis of the graph Laplacian $L$, the discrete counterpart of the continuum Laplacian $\nabla^2$ \cite{Burioni1996, Binney}. For a ferromagnetic interaction graph, $L=D-J$, where $J$ is the nonnegative interaction matrix and $D$ is the diagonal strength matrix, $D_{ii}=\sum_j J_{ij}$. The eigenvalues $\lambda_\alpha$ and eigenvectors $\ket{\lambda_\alpha}$ of $L$ then provide the graph analogue of Fourier modes, encoding the connectivity and geometry of the underlying lattice \cite{LRG, Iannelli2025, Villegas2025}. More than an analogy, this operator is tied to \eqref{Ising} by an exact identity: since $\sigma_i^2=1$, $\mathcal{H}(\vec\sigma)=\tfrac{1}{2}\,\vec\sigma^{\transp}\slapl\,\vec\sigma-\tfrac{1}{2}\sum_i\abs{D}_{ii},$ where $\slapl=\abs{D}-J$ is the signed Laplacian, $\abs{D}_{ii}=\sum_j\abs{J_{ij}}$, and $\slapl=L$ for ferromagnetic couplings. The subtracted term depends only on the
interaction graph, so minimizing the Ising energy over $\{\pm1\}^N$ is equivalent to minimizing the Laplacian quadratic form on the hypercube.

By applying the sign function to these eigenvectors~\footnote{We note that, strictly speaking, classical Chladni patterns refer to the nodal lines of vibrational eigenmodes, whereas here a Chladni state denotes the full binary spin configuration obtained by component-wise binarization of a Laplacian eigenvector.}, we obtain binary spin configurations that we call \emph{Chladni states}, $\sigma_i^{(\alpha)}=\sign\psi_i^{(\alpha)}$, where $\psi_i^{(\alpha)}$ is the $i$-th component of the Laplacian eigenvector
$\psi_i^{(\alpha)}=\braket{i|\lambda_\alpha}$. Their energies then follow from the identity above, evaluated on the binarized eigenvector
rather than on the eigenvector itself: whether they still track the spectral ordering, or remain dynamically stable, is a nontrivial property
of this nonlinear map, to be established rather than assumed. In this view, Chladni states provide a graph-spectral dictionary of binary spin
patterns generated by the interaction geometry, much like plane-wave decomposition does for continuum fields.

\begin{figure}[hbtp]
    \centering
    \includegraphics[width=1.0\columnwidth]{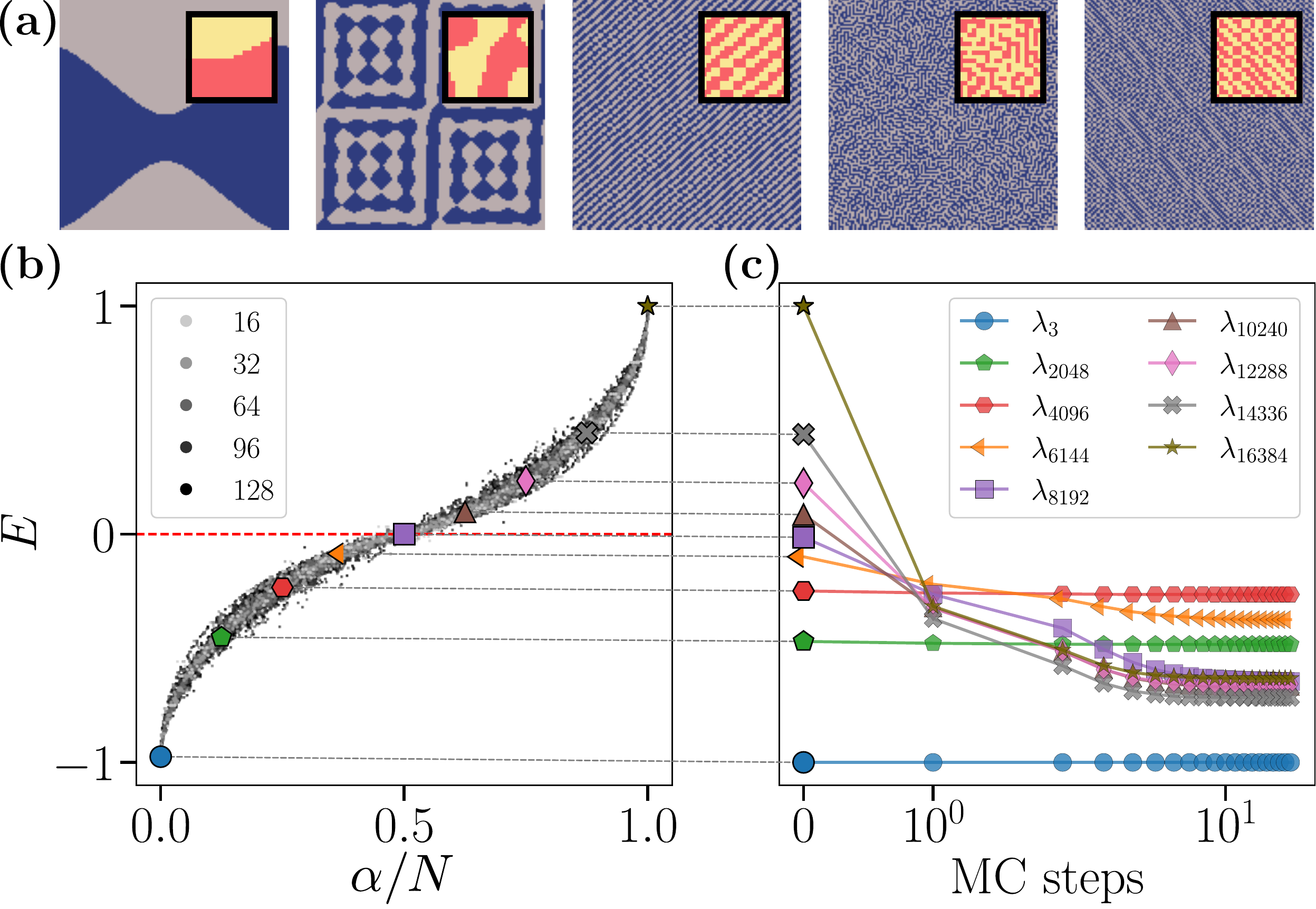}
    \caption{\textbf{Chladni states in a ferromagnetic square lattice.}
    \textbf{(a)} Selected Chladni states obtained by binarizing Laplacian eigenvectors of a square lattice with $N=128\times128=16384$ sites. From left to right: $\alpha=(2,~256,~4096,~8192,~12288)$. Insets: Zoomed patterns.
    \textbf{(b)} Ising energy of the Chladni states versus the normalized eigenmode index $\alpha/N$, for several system sizes ($N=L^2$; see legend). The curves collapse across sizes and exhibit an inflection point at $\alpha/N=1/2$, separating the negative- and positive-energy sectors. The red dashed line marks zero energy.
    \textbf{(c)} Zero-temperature Monte Carlo evolution initialized from selected Chladni states for $L=128$. Negative-energy states remain stable in time, whereas positive-energy states relax toward lower-energy configurations. Periodic boundary conditions are used throughout.}
    \label{fig:panel_1}
\end{figure}

Figure~\ref{fig:panel_1}(a) shows selected Chladni states for a 2D square lattice (additional examples for other lattices are reported in the SM~\cite{SM}). On the periodic square lattice, Laplacian eigenvalues are often degenerate. Numerical diagonalization returns one particular real orthonormal basis of each degenerate eigenspace; the analytic Fourier, or standing-wave, modes span the same eigenspace and are an equally legitimate choice. Since binarization is nonlinear, orthogonal rotations of the eigenbasis can produce visually different Chladni states while leaving the underlying eigenspace invariant. On the periodic square lattice, for instance, the low-index Chladni states are the striped configurations labelled by the winding numbers $(a,b)$ of the mode wavevector $\mathbf k=2\pi(a,b)/L$, that is, the long-lived states of the zero-temperature kinetic Ising model, whose occurrence probabilities are related to critical percolation~\cite{Redner2001,Redner2009,Olejarz2012}. Numerically diagonalized eigenvectors, being generic rotations within the same degenerate eigenspace, typically binarize into diagonal or blob-like mixtures of these stripes.

\FigRef{fig:panel_1}(b) shows the associated Ising energy of each Chladni state as a function of the normalized eigenmode index. The resulting curve exhibits a characteristic S-shape that collapses across system sizes, while remaining sensitive to the underlying lattice geometry (see SM~\cite{SM}). In particular, Chladni states associated with eigenvectors in the lower half of the spectrum have negative Ising energies. These low-index modes correspond to large-scale patterns with broad domains, so that only a small fraction of bonds is unsatisfied. This intuition has a precise geometric counterpart. Since $\sigma_i\sigma_j=-1$ precisely on the bonds cut by the nodal set of $\ket{\lambda_\alpha}$, counting satisfied and cut bonds gives $E/N=-z/2+2W/N$, with $z$ the coordination number and $W$ the number of cut bonds -- so the energy is, up to an affine shift, the nodal perimeter of the eigenvector. The S-shaped curve of \FigRef{fig:panel_1}(b) is therefore the growth of this nodal-line density with mode index. A related but distinct constraint bounds how many separate domains that perimeter can enclose: the discrete analogue of Courant's theorem limits the number of nodal domains of the mode indexed by \(\alpha\) to at most \(\alpha+1\)~\cite{Sudhir2017}.

To test their stability, we simulate a kinetically constrained zero-temperature Ising dynamics (see~\cite{Godreche2005} for further details), using random single-spin updates. The microscopic rule is as follows: at each elementary step, a spin is selected uniformly at random, and spin flips are accepted only when they lower the energy, while moves with $\Delta E=0$ are rejected~\footnote{The treatment of $\Delta E=0$ moves is important: standard Glauber dynamics can smooth domain walls through zero-energy flips, making some patterns unstable. Refs.~\cite{Redner2001,Redner2009,Olejarz2012} allow $\Delta E=0$ moves; forbidding them enlarges the frozen family.}. One Monte Carlo sweep corresponds to $N$ attempted spin updates. We initialize the system in selected Chladni states and follow its \gls{mc} evolution, as shown in \FigRef{fig:panel_1}(c). Low-index Chladni states with negative energy remain dynamically trapped, whereas higher-energy states relax toward lower-energy configurations. For $0<T<T_c$, thermal fluctuations give these metastable patterns a finite lifetime and ultimately restore equilibration toward the ordered ferromagnetic phase.

The same behavior is observed in other two-dimensional geometries, including triangular, hexagonal, and two-scale square-octagon lattices inspired by real materials~\cite{Falsi2024} (see SM~\cite{SM}). Each geometry generates a distinct family of Chladni states and a corresponding energy profile. We therefore take the Laplacian eigenvectors $\ket{\lambda_\alpha}$ as the basis in which frozen spin configurations are decomposed. 
We index the eigenmodes increasingly as
$\alpha=0,\ldots,N-1$. The reconstruction using the first $K$ Laplacian eigenvectors is then defined as
\begin{equation}
|m^{(K)}\rangle =
\operatorname{sign}\!\left[
\sum_{\alpha=0}^{K-1}
\langle \lambda_\alpha | m\rangle
|\lambda_\alpha\rangle
\right].
\label{eq:reconstruction}
\end{equation}
where the sign is applied independently to each component of the vector. We refer to this reconstruction procedure as \emph{\acrfull{gsd}}. Although the Laplacian eigenvectors form a complete basis, the relevant question is how much of a metastable pattern survives when the reconstruction is restricted to a few low-lying modes and then binarized.

Given a spin configuration $\ket{m}$ and its reconstruction $\ket{m^{(K)}}$ from the first $K$ Laplacian modes, we quantify reconstructability as
\begin{equation}
    R(K)=\frac{1}{N}\sum_{i=1}^{N}
    \delta_{m_i,m_i^{(K)}} ,
    \label{eq:reconstructability}
\end{equation}
where $m_i$ and $m_i^{(K)}$ are the $i$-th components of the original and reconstructed spin configurations, respectively. Thus, $R(K)$ is the fraction of correctly reconstructed spins. A sharp increase of $R(K)$ at a given $K$ therefore identifies the spectral band that carries most of the structure of the pattern. A linear counterpart of $R(K)$ is the projected weight
$C(K)=\frac{1}{N}\sum_{\alpha=0}^{K-1}|\langle m|\lambda_\alpha\rangle|^2$, which identifies the same spectral bands but saturates more slowly,
since it measures amplitude rather than sign (see SM~\cite{SM}).

\section{GSD in frustrated antiferromagnetic systems}
\begin{figure}[hbtp]
    \centering
\includegraphics[width=1.0\columnwidth]{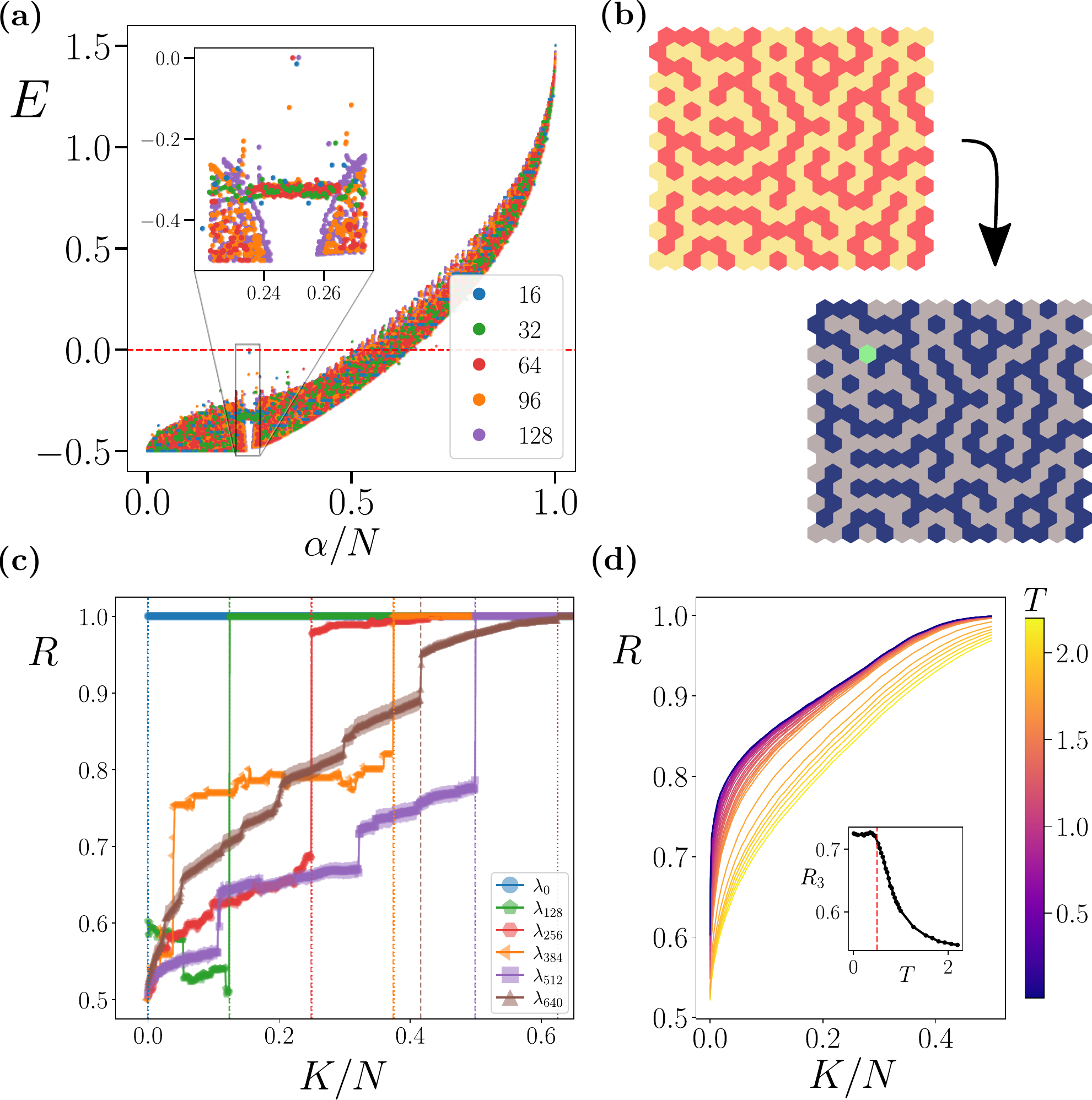}
        \caption{\textbf{GSD in a frustrated triangular antiferromagnet.}
        \textbf{(a)} Ising energy of Chladni states versus normalized eigenmode index for antiferromagnetic triangular lattices of different sizes ($N=L^2$; see legend). The curves collapse across system sizes. The red dashed line marks zero energy.
        \textbf{(b)} Metastable state obtained from a random initial condition after long-time zero-temperature dynamics (top) and its reconstruction via \gls{gsd} (bottom) using \eqref{eq:reconstruction} with the first $K=108$ signed-Laplacian eigenvectors ($N=256$). A mismatched spin is highlighted in green.
        \textbf{(c)} Reconstructability index $R(K)$ as a function of the normalized number of retained modes, $K/N$, for configurations initialized in selected Chladni states at $T=0$ (initial condition reported in the legend; $t=40N$ \gls{mc} steps; \(n_{\rm avg} = 20\) independent update histories).
        \textbf{(d)} $R(K)$ for long-time states obtained from random initial conditions at different temperatures ($t=60N$ \gls{mc} steps; \(T\) in colorbar; Each curve is averaged over \(500\) independent thermal histories). Inset: $R$ as a function of $T$ using the first three signed-Laplacian modes. The change around $T\simeq0.5$ marks the crossover into the low-temperature regime, where $\xi\gtrsim L$; above it, more modes are required to reconstruct the spin pattern. Parameters: $N=1024$.}
    \label{fig:panel_2}
\end{figure}

We apply GSD to the paradigmatic case of the triangular-lattice antiferromagnet, where geometric frustration prevents the system from achieving conventional global long-range order \cite{Drisko2017, Ramirez2003}. The relevant modes are those of the signed Laplacian $\slapl=\abs{D}-J$, which for a uniform antiferromagnet reduces to $\slapl=z\,I+\adj$, with $\adj$ the adjacency matrix and $z=6$ the coordination number; the index $\alpha$ in \FigRef{fig:panel_2}(a) orders its eigenvalues $\bar\lambda_\alpha=z+a_\alpha$ increasingly, $a_\alpha$ being the adjacency eigenvalues. The resulting Chladni-state energy profile is shown in \FigRef{fig:panel_2}(a).

In particular, low-index modes correspond to metastable configurations at $T=0$ (see SM~\cite{SM}). The asymmetry of the curve reflects the fact that, on the triangular lattice, the antiferromagnetic problem is not equivalent to the ferromagnetic one by a simple sublattice spin flip, in contrast to bipartite lattices such as the square lattice. The narrow feature in \FigRef{fig:panel_2}(a) is the saddle-point
(van Hove) level $a=-2$ of the triangular-lattice band. For even $L$, it has degeneracy $g=3L-3$; the corresponding block is centered at
normalized spectral rank $1/4$, with exact relative width $g/N=3/L-3/L^2$. The eigenspace is generated by the three reciprocal-space line families $k_1=\pi$, $k_2=\pi$, and $k_1-k_2=\pi$. Any real superposition restricted to one family factorizes into an alternating
longitudinal component and an arbitrary transverse profile. For non-nodal representatives, binarization gives $\sigma_{n_1n_2}=(-1)^{n_1}\tau_{n_2}$, or a symmetry-related form. These non-nodal family-resolved states satisfy $\sum_{j\in\partial i}\sigma_i\sigma_j=-2$ at every site, and are therefore exact ground states ($E_{\rm AF}/N=-1$) with no free spins, strictly frozen under single-spin dynamics since $\Delta E_i=+4$ for every flip. Collective chain flips cost zero and generate the $2^L$ members of each family~\cite{Diep2013}. Generic vectors in the full degenerate eigenspace typically mix the three families; after binarization they need not retain the factorized form, be ground states, or remain dynamically frozen, producing the residual scatter in \FigRef{fig:panel_2}(a). Ground states at the same energy may instead contain free spins and, when zero-energy flips are allowed, move within the degenerate manifold. On the square lattice, the analogous saddle level lies at $a=0$, has degeneracy $2L-2$, and yields $\sum_{j\in\partial i}\sigma_i\sigma_j=0$ for the corresponding family-resolved states, accounting for the zero-energy crossing at $\alpha/N=1/2$ in \FigRef{fig:panel_1}(b) (see also SM~\cite{SM}). At a small finite temperature, the absence of a fully ordered phase allows the dynamics to explore the degenerate low-energy landscape \cite{Shokef2011}, and the long-time state can be expressed as a combination of low-lying Laplacian modes. Initial conditions with negative energy leave a clear memory imprint on the final metastable configuration: the dynamics does not leave the region of the landscape in which it started (see SM~\cite{SM}). \FigRef{fig:panel_2}(b) shows a metastable state obtained from a random initial condition at $T=0$ together with its reconstruction via \gls{gsd}.

Figure~\ref{fig:panel_2}(c) shows the reconstructability index $R(K)$ for a triangular antiferromagnetic lattice initialized in selected Chladni states at $T=0$, while \FigRef{fig:panel_2}(d) reports $R(K)$ for long-time states obtained from random initial conditions at different temperatures. A clear change occurs around $T\simeq0.5$. The nearest-neighbour triangular Ising antiferromagnet
with $S=1/2$ has no finite-temperature transition~\cite{Wannier1950,Chung2006}: for the system sizes and observation times considered here, $T\simeq0.5$ marks a finite-size dynamical crossover into the
low-temperature regime, where the energy density approaches the Wannier value and the correlation length $\xi$ becomes comparable to the linear size $L$. Below this crossover a few long-wavelength modes suffice to describe the configuration; above it, with $\xi<L$, increasingly many modes are required. Related higher-spin or continuous-spin triangular models can instead exhibit finite-temperature transitions~\cite{Yamada1995,Chung2006,Kawamura1984}. In both regimes, GSD acts as a spectral filter: by restricting the reconstruction to selected Laplacian modes, it quantifies which spectral bands encode the dominant structure of a metastable configuration.
\newpage

\section{GSD in Spin Glasses}
We now turn to \glspl{sg} \cite{Edwards1975}, a paradigmatic class of non-ergodic systems in which competing ferromagnetic and antiferromagnetic bonds generate frustration, multiple metastable states, and slow relaxation in a rugged energy landscape \cite{Bray1987, Castellani2005, Mezard1984}. We consider the random-bond Ising model, where $p$ corresponds to the fraction of negative links. Frustration is naturally accounted for by the signed Laplacian $\slapl$ introduced above, which remains positive-semidefinite for arbitrary bond signs and provides the graph-spectral basis used for GSD~\cite{Iannelli2025,Liu2015}.

\begin{figure}[hbtp]
    \centering
    \includegraphics[width=\columnwidth]{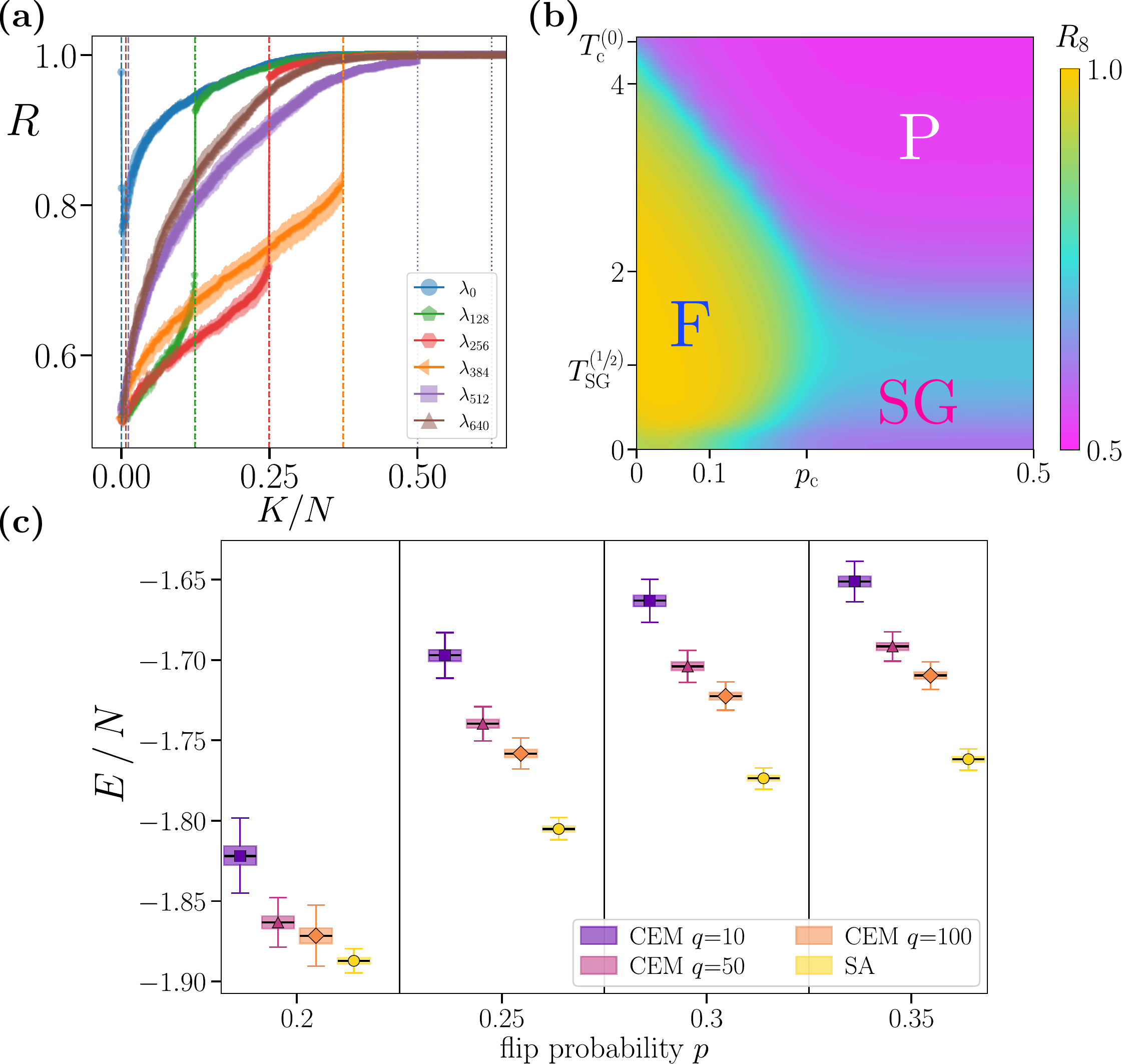}
    \caption{\textbf{GSD in random-bond SGs.}
    \textbf{(a)} Reconstructability, $R(K)$, versus normalized number of retained modes $K/N$, for selected Chladni-state initial conditions at $T=0$ in a 3D cubic \gls{sg} ($t=30N$ \gls{mc} steps, $p=0.25$, $N=4096$). Sharp increases reveal memory of the initial spectral sector.
    \textbf{(b)} Phase diagram as a function of the fraction of negative bonds \(p\) and temperature \(T\) for a 3D cubic \gls{sg} initialized at random ($N=4096$). The color code shows the reconstructability using the first eight signed-Laplacian modes, where the three regimes (ferromagnetic, paramagnetic, and spin-glass) appear as distinct levels of $R_8$, with boundaries consistent with the known phase diagram of the model. Each point is averaged over $n_{\rm dis}=1000$ independent disorder realizations; for each realization, one independent thermal history is simulated.
    \textbf{(c)} Low-dimensional spectral search for low-energy SG configurations. For a 3D random-bond Ising SG with $L=30$ and $N=L^3$, we compare simulated annealing with a CEM optimization restricted to configurations obtained by binarizing fields spanned by the first \(q\) modes. Energies are reported as energy densities, $E/N$, so the comparison is normalized by system size. The restricted CEM search reaches energy densities comparable to the simulated-annealing benchmark, showing that low-energy configurations in frustrated landscapes can be approximated in a low-dimensional signed-Laplacian space. Averages are taken over 800 independent disorder realizations, with 10 thermal histories per realization.}
    \label{fig:panel_3}
\end{figure}

\FigRef{fig:panel_3}(a) shows the reconstructability index $R(K)$ for a random-bond Ising model on a 3D cubic lattice at $T=0$, initialized in different Chladni states. The sharp increase of $R(K)$ near the initialization mode shows that the memory of the initial spectral sector is retained by the subsequent dynamics, as expected for trapping in long-lived spin-glass basins. As in the antiferromagnetic case, the number of modes needed to reconstruct the configuration grows sharply on entering the paramagnetic regime [\FigRef{fig:panel_3}(b)], so that GSD also detects the loss of mesoscopic order in short-range frustrated systems.

Finally, we test whether the low-lying signed-Laplacian modes can approximate low-energy minima in a spin-glass energy landscape. As a benchmark, we consider a 3D random-bond Ising spin glass with $L=30$ and compare with simulated annealing. Instead of searching directly in the full configuration space $\{\pm1\}^N$, we restrict the search to configurations of the form $\vec{\sigma}(\vec c)=\mathrm{sign}(V_q\vec c)$, where $V_q$ contains the first $q$ eigenvectors of the signed Laplacian and the coefficients $\vec c$ are optimized with a cross-entropy method~\cite{deBoer2005CEM} (CEM; see SM~\cite{SM}). The cross-entropy method is a natural choice for this reduced search because it is a gradient-free population optimizer, well-suited to the non-smooth map $\vec c\mapsto \mathrm{sign}(V_q\vec c)$ and to rugged energy landscapes with many local minima. As shown in \FigRef{fig:panel_3}(c), this restricted spectral search reaches energies comparable to the simulated-annealing benchmark. A small number of Chladni modes therefore provides an effective approximation space for low-energy configurations in frustrated systems with many metastable states.

\section{Encoding patterns in the spectrum of a geometry}
Chladni states describe which spin patterns a given interaction geometry can sustain. The question can be turned around: which arbitrary patterns does the spectrum of that geometry encode compactly? Any field defined on the lattice sites, not necessarily a spin configuration, can be projected onto the graph modes, and the number of low-lying modes needed to reproduce it measures how well its spatial correlations align with those favored by the geometry. The interaction network thus defines a spectral dictionary of its own.

On regular translation-invariant lattices this dictionary is naturally connected to standard Fourier-, cosine-, and Walsh-type representations~\cite{Ahmed1974DCT,Walsh1923,Shuman2013}. Indeed, on such lattices Laplacian eigenvectors reduce to graph-Fourier modes, and the component-wise binarization of sinusoidal modes produces telegraph-like binary patterns reminiscent of Walsh functions. Yet, \gls{gsd} is not a Walsh--Hadamard transform: there, the binary basis is fixed by a dyadic construction and an associated ordering, whereas here it is determined by the
Laplacian of the chosen interaction graph, with binary patterns arising from a nonlinear sign projection of real-valued eigenvectors -- a construction that survives precisely where no dyadic or Fourier ordering is available, as in generic signed, frustrated, or irregular graphs. It also differs from Laplacian eigenmaps~\cite{Belkin2003}, where the graph is built from the data themselves: here the basis is fixed a priori by the interaction network, and the data are projected onto physically defined modes. In this way, \gls{gsd} provides a compact and interpretable graph-spectral representation of frozen spin, dipolar, or image-like patterns, potentially relevant to frustrated systems and materials with microstructural disorder~\cite{Xin2024, Simonov2020, Ramesh2019, Nahas2020}.

\begin{figure}[hbtp]
    \centering
    \includegraphics[width=\columnwidth]{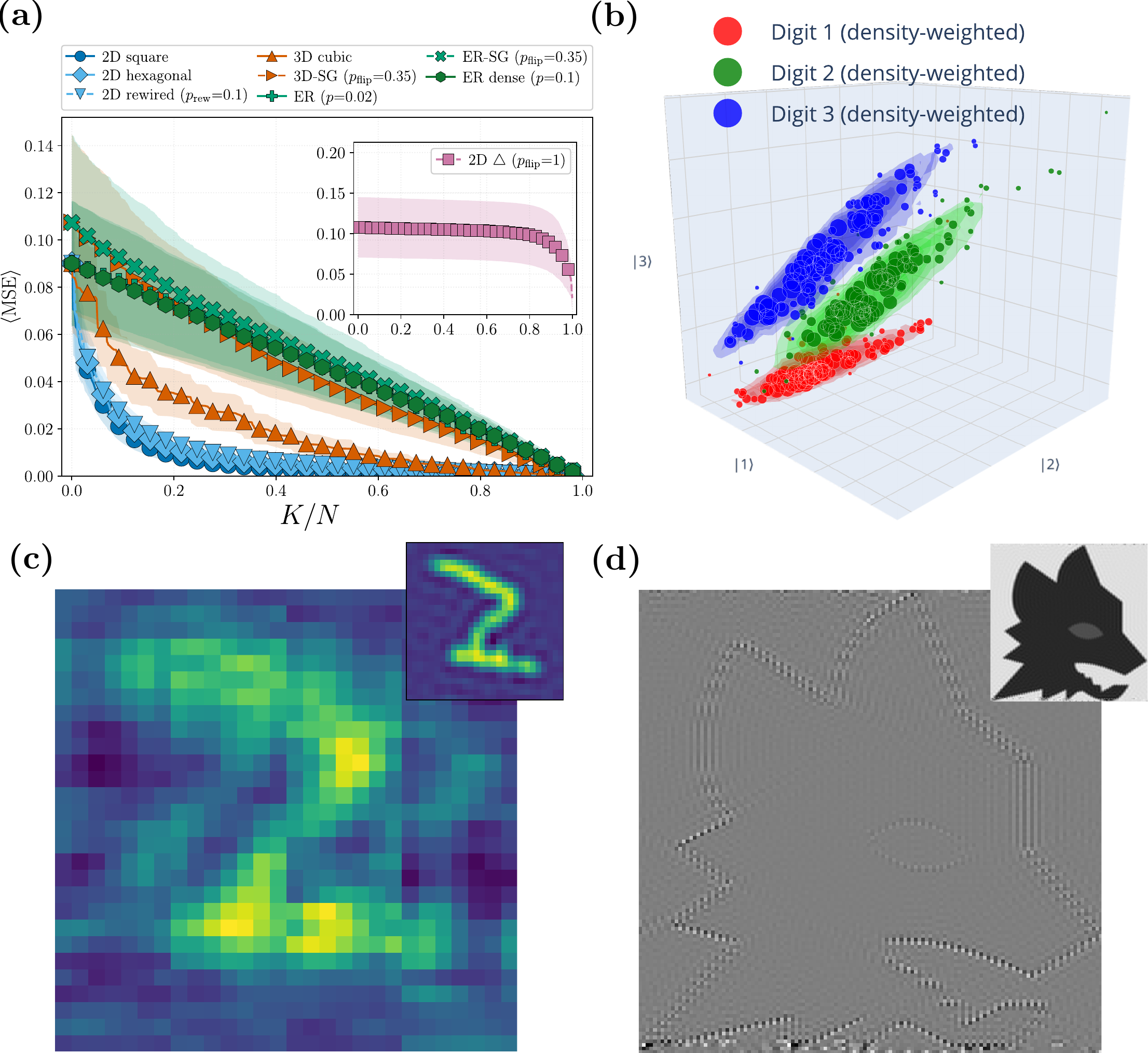}
    \caption{\textbf{Patterns encoded by the spectrum of different geometries.}
    \textbf{(a)} Average mean-squared error (MSE) between original and reconstructed digits as a function of the number of retained Laplacian modes, for different graph topologies, using $5\cdot10^2$ randomly selected MNIST images.
    \textbf{(b)} Projection of MNIST digit images onto the first three Laplacian eigenvectors of a signed 2D square lattice with $p=0.08$ and $L=28$.
    \textbf{(c)} Reconstruction of a digit using the low-lying $K=200$ modes of a three-dimensional lattice basis ($N=14\times8\times7=784$), compared with the reconstruction obtained from a 2D square-lattice basis ($N=28\times28=784$) in the inset. Different graph geometries select different spatial correlations and therefore encode the same pattern with different efficiency.
    \textbf{(d)} Reconstruction of a binary image ($128\times128$) using the first $K=6000$ modes of a 2D antiferromagnetic triangular-lattice basis, compared with the reconstruction obtained from a 2D ferromagnetic triangular-lattice basis in the inset. Even at fixed dimension, interaction signs and lattice symmetries modify the spectral filter and the resulting reconstructed pattern.}
    \label{fig:mnist}
\end{figure}

We use handwritten digits as a neutral probe, treating each image as a field on the lattice sites. \FigRef{fig:mnist}(a) shows the average mean-squared error, $\mathrm{MSE}=N^{-1}\sum_i (x_i-x_i^{(K)})^2$, between original images $x$ and reconstructions $x^{(K)}$, as a function of the number $K$ of retained modes for different graph topologies. The number of modes needed to reach a given error differs between the best- and the worst-matched geometry (see also SM~\cite{SM}). The encoding and decoding are therefore fixed analytically by the interaction graph and require no training, providing a geometry-defined analogue of the bottleneck representation of an autoencoder~\cite{Hinton2006}. \FigRef{fig:mnist}(b) shows the resulting three-dimensional embedding of the digit classes obtained from the first three non-trivial modes of a signed square lattice.

Figures~\ref {fig:mnist}(c,d) show that both the quality and the character of the reconstruction depend on the interaction geometry and its spectral ordering. Different graph geometries assign low spectral cost to different spatial correlations, so that the same pattern can be encoded more compactly by one geometry than by another. Even at fixed dimension, changing the interaction signs modifies the spectral ordering and therefore which spatial correlations occupy the low-lying spectral sector [\FigRef{fig:mnist}(d)]. Interaction sign provides a particularly transparent connection with the Chladni construction. For the triangular lattice, the same change that separates the ferromagnetic and antiferromagnetic Chladni-energy profiles (SM~\cite{SM} and \FigRef{fig:panel_2}) also changes the truncated reconstructions of \FigRef{fig:mnist}(d). The two readings are not equivalent, however. Compact encoding is a linear property of the overlap of a pattern with a low-dimensional spectral subspace, whereas freezing emerges only after the nonlinear sign projection and the subsequent spin dynamics. Yet both are governed by the same geometric prior: the spectrum determines which spatial correlations are spectrally simple, while its binarized modes define a distinguished family of candidate frozen states. In this sense, the interaction geometry defines not only the energetics of spin patterns, but also a geometry-dependent notion of representational complexity.

\section{Outlook}
Our results introduce a graph-spectral framework for analyzing metastability in discrete spin systems. By binarizing Laplacian eigenmodes, we obtain Chladni states: spin configurations whose ferromagnetic energy is, up to a constant, the nodal perimeter of the eigenvector, and whose dynamical stability is systematically correlated with their position in the spectrum of the interaction network. For classical Ising ferromagnets, low-lying Chladni states provide a distinguished family of configurations that are locally stable against zero-temperature single-spin dynamics. In frustrated antiferromagnets and \glspl{sg}, the same decomposition identifies the spectral sectors that retain memory of metastable basins.

On highly symmetric lattices, spectral degeneracies make the family of Chladni states larger than the spectrum alone would suggest. A degenerate Laplacian eigenspace admits several equivalent real orthonormal bases, whose nonlinear binarizations can give different binary representatives of the same spectral subspace. Thus, individual Chladni patterns should be interpreted together with the eigenspaces from which they originate, while subspace-based quantities, such as truncated reconstructions, give a description that does not depend on the particular eigenbasis chosen.

The triangular antiferromagnet provides a striking consequence of this spectral degeneracy. Within the highly degenerate van Hove eigenspace, the three symmetry-resolved line families generate, after binarization, a distinguished subset of exact ground states with no free spins, whereas generic mixtures of the same modes need not be ground states
or dynamically frozen. Nonlinear binarization therefore turns different directions within the same linear spectral subspace into physically inequivalent spin configurations.

Taken together, these results suggest using \gls{gsd} as a reduced description of multistate energy landscapes. As a diagnostic, it reveals which spectral bands carry the structure of a frozen configuration and tracks the persistence of memory and mesoscopic order where no simple scalar order parameter is available. As a search space, a low-dimensional set of signed-Laplacian modes is sufficient, in random-bond spin glasses,
to reach energies comparable to simulated annealing, showing that low-energy configurations can be efficiently approximated within a spectral subspace fixed by the interaction network. As a representation, the same basis acts as an analytical, geometry-defined analogue of an autoencoder, in which the interaction graph fixes the encoding and decoding without training. The same interaction-defined spectrum thus governs frozen-state
structure, low-energy search, and pattern representation. This perspective may be useful for interpreting frozen magnetic or dipolar textures in frustrated materials, and more broadly for understanding how memory-like structures can emerge from the spectral geometry of interacting systems.

\emph{\textbf{Acknowledgments--}}We thank A. Gabrielli for useful discussions and comments. P.V. acknowledges the Spanish Ministry and Agencia Estatal de Investigaci\'on (AEI), MICIN/AEI/10.13039/501100011033, for financial support through Project PID2023-149174NB-I00, funded also by ERDF/EU.\vspace{-2mm}

\emph{\textbf{Data availability--}}The data that support the findings of this article are available from the authors upon reasonable request. The code used to generate and analyze them is openly available in Ref.~\cite{lrgsglib}.\vspace{-2mm}

%

\clearpage
\pagestyle{empty}
\includepdf[pages={{},-},pagecommand={\thispagestyle{empty}}]{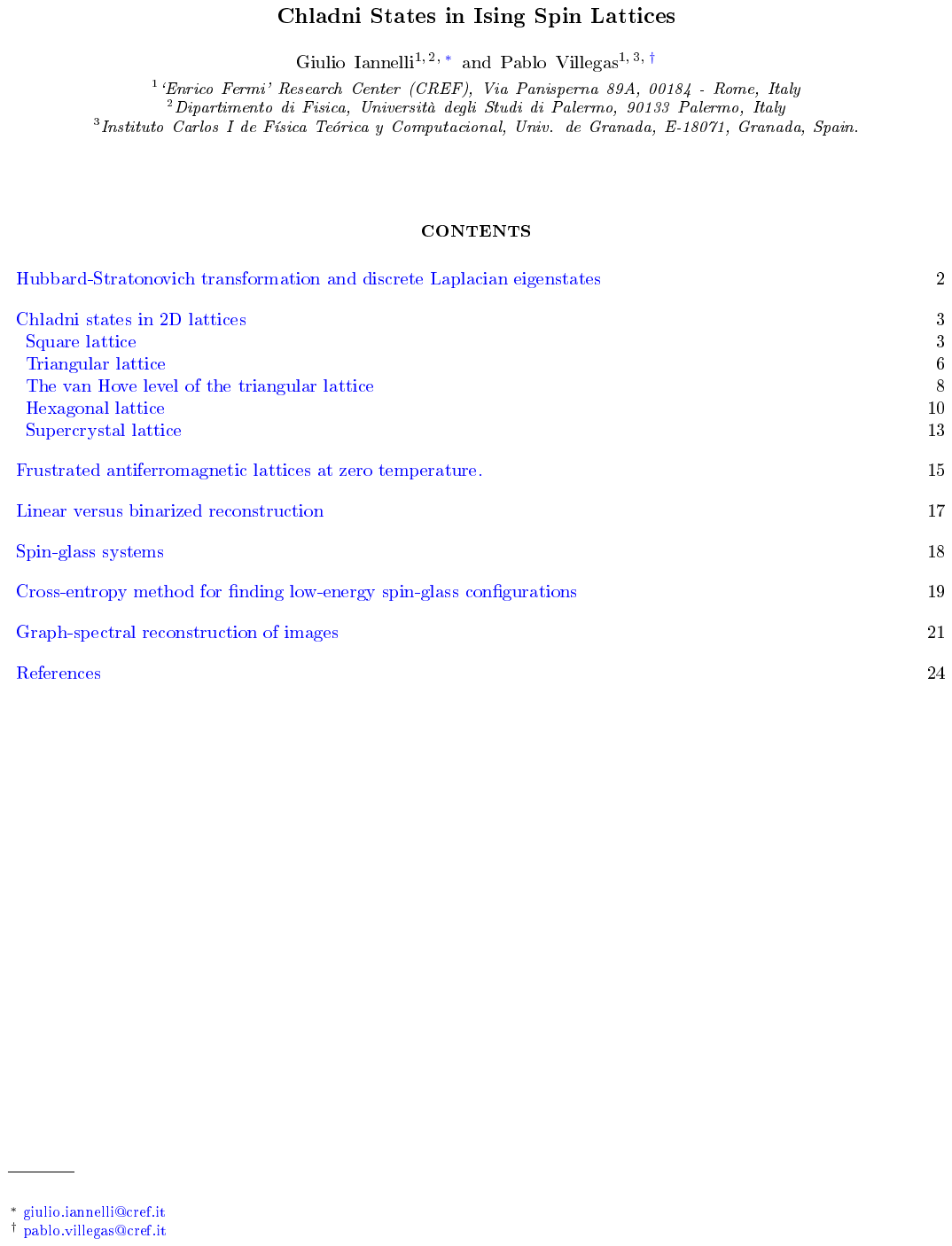}

\end{document}